\documentclass[12pt]{article}

\usepackage{latexsym} % Gets \Box etc
\usepackage{amssymb} % \gtrsim, \geqslant, etc etc: see amsguide.ps
\usepackage{graphicx} 

\makeatletter

\def\section{
\setcounter{equation}{0}        % Reset eqn numbers at start of section
\@startsection {section}{1}{\z@}{-3.5ex plus -1ex minus 
 -.2ex}{2.3ex plus .2ex}{\large\bf}}

\def\subsection{\@startsection{subsection}{2}{\z@}{-3.25ex plus -1ex minus 
 -.2ex}{1.5ex plus .2ex}{\normalsize\bf}}

\def\subsubsection{\@startsection{subsubsection}{3}{\z@}{-3.25ex plus
 -1ex minus -.2ex}{1.5ex plus .2ex}{\normalsize}}

\makeatother   %\catcode`\@=12

\newcommand{\Dash}{\boldmath $-$}  % for item symbol

\newcommand{\al}{\alpha}
\newcommand{\be}{\beta}

\newcommand{\De}{\Delta}

\newcommand{\eps}{\epsilon}

\newcommand{\ka}{\kappa}

\newcommand{\beq}{\begin{equation}}
\newcommand{\eeq}{\end{equation}}
\newcommand{\ba}{\begin{array}}
\newcommand{\ea}{\end{array}}
\newcommand{\bea}{\begin{eqnarray}}
\newcommand{\eea}{\end{eqnarray}}
\newcommand{\bi}{\begin{itemize}}  %\setlength{\itemsep}{0\parsep}}
\newcommand{\ei}{\end{itemize}}
\newcommand{\ben}{\begin{enumerate}} %\setlength{\itemsep}{0\parsep}}
\newcommand{\een}{\end{enumerate}}
\newcommand{\bc}{\begin{center}}
\newcommand{\ec}{\end{center}}

\renewcommand{\>}{\rangle} % LaTeX: \> already defined
\newcommand{\txt}{\textstyle}

\newcommand{\half} {{\txt \frac{1}{2}}}
\newcommand{\third}{{\txt \frac{1}{3}}}

\newcommand{\twothirds}{{\txt \frac{2}{3}}}
\newcommand{\MeV}{{\rm MeV}} 

\newlength{\vpad}
\begin{document}

\title{\bf Color superconductivity and the strange quark}

%\classification{12.38.-t, 26.60.+c, 25.75.Nq}
%\keywords      {dense quark matter, color superconductivity}
\medskip

\author{Mark Alford \\[1ex]
 Physics Department\\ Washington University \\
Saint Louis, MO 63130\\ USA
}

\date{Nov 22, 2005}

\begin{titlepage}
\maketitle

\begin{abstract}
At ultra-high density, matter is expected to form a degenerate
Fermi gas of quarks in which there is
a condensate of Cooper pairs of quarks near the Fermi surface:
color superconductivity. In these proceedings I review some
of the underlying physics, and discuss outstanding questions
about the phase structure of ultra-dense quark matter.
% possible signatures by which color superconducting quark matter might
% be detected in compact stars.
\end{abstract}

\end{titlepage}

\section{Introduction}
\label{sec:intro}

The exploration of the phase diagram of matter at ultra-high temperature
or density is an area of great interest and activity, both on the
experimental and theoretical fronts. Heavy-ion colliders such as
the SPS at CERN and RHIC at Brookhaven have probed the high-temperature
region, searching for the transition to deconfined quark matter.
In this paper we discuss a different part of
the phase diagram, the low-temperature high-density region. 
Here there are
as yet no experimental constraints, but we expect to find
phases characterized by Cooper pairing of quarks, i.e.
 color superconductivity, driven by
the Bardeen-Cooper-Schrieffer (BCS)
\cite{BCS} mechanism. The BCS mechanism operates when there exists
an attractive interaction between fermions at a Fermi surface.
The QCD quark-quark interaction is strong, and is attractive
in many channels, so we expect cold dense quark matter to {\em generically}
exhibit color superconductivity.
Moreover, quarks, unlike electrons, have color and flavor as well as spin
degrees of freedom, so many different patterns of pairing are possible.
This leads us to expect a rich phase structure
in matter beyond nuclear density.

Calculations using a variety of methods agree that
at sufficiently
high density, the favored phase is color-flavor-locked (CFL)
color-superconducting quark matter \cite{CFL} (for reviews,
see Ref.~\cite{Reviews}). 
However, there is still uncertainty over the nature of the next phase down
in density. Recent
work \cite{gCFL} suggests that when the density drops low enough so that
the mass of the strange quark can no longer be neglected, there
is a continuous phase transition from the CFL phase to a new
gapless CFL (gCFL) phase, which could lead to observable consequences
if it occurred in the cores of neutron stars \cite{Alford:2004zr}.
However, it now appears that some of the gluons
in the gCFL phase have imaginary
Meissner masses, indicating an instability towards 
an unknown lower-energy phase
\cite{Huang:2004am,Casalbuoni:2004tb,Giannakis:2004pf,Fukushima:2005cm}. 
The nature of this phase is still unclear, although the
crystalline ``LOFF'' phase is a strong candidate 
\cite{Alford:2000ze,Giannakis:2004pf}.

% This question is not a purely theoretical one.
% Color superconducting quark matter may occur naturally in the
% universe, in the cold dense cores of compact (``neutron'') stars,
% where densities are above nuclear density, and temperatures are of the
% order of tens of {\rm keV}.  (It might conceivably be possible to
% create it in future low-energy heavy ion colliders, such as the Compressed
% Baryonic Matter facility at GSI Darmstadt.)  Up to now, most work on
% signatures has focussed on properties of color superconducting quark
% matter that would affect observable features of compact stars, and I
% will discuss some of these below.

\section{Review of color superconductivity}

\subsection{The phase diagram of quark matter}

\begin{figure}[t]
\parbox{0.44\hsize}{
\begin{center} \underline{Conjectured form} \end{center}
 \includegraphics[width=\hsize]{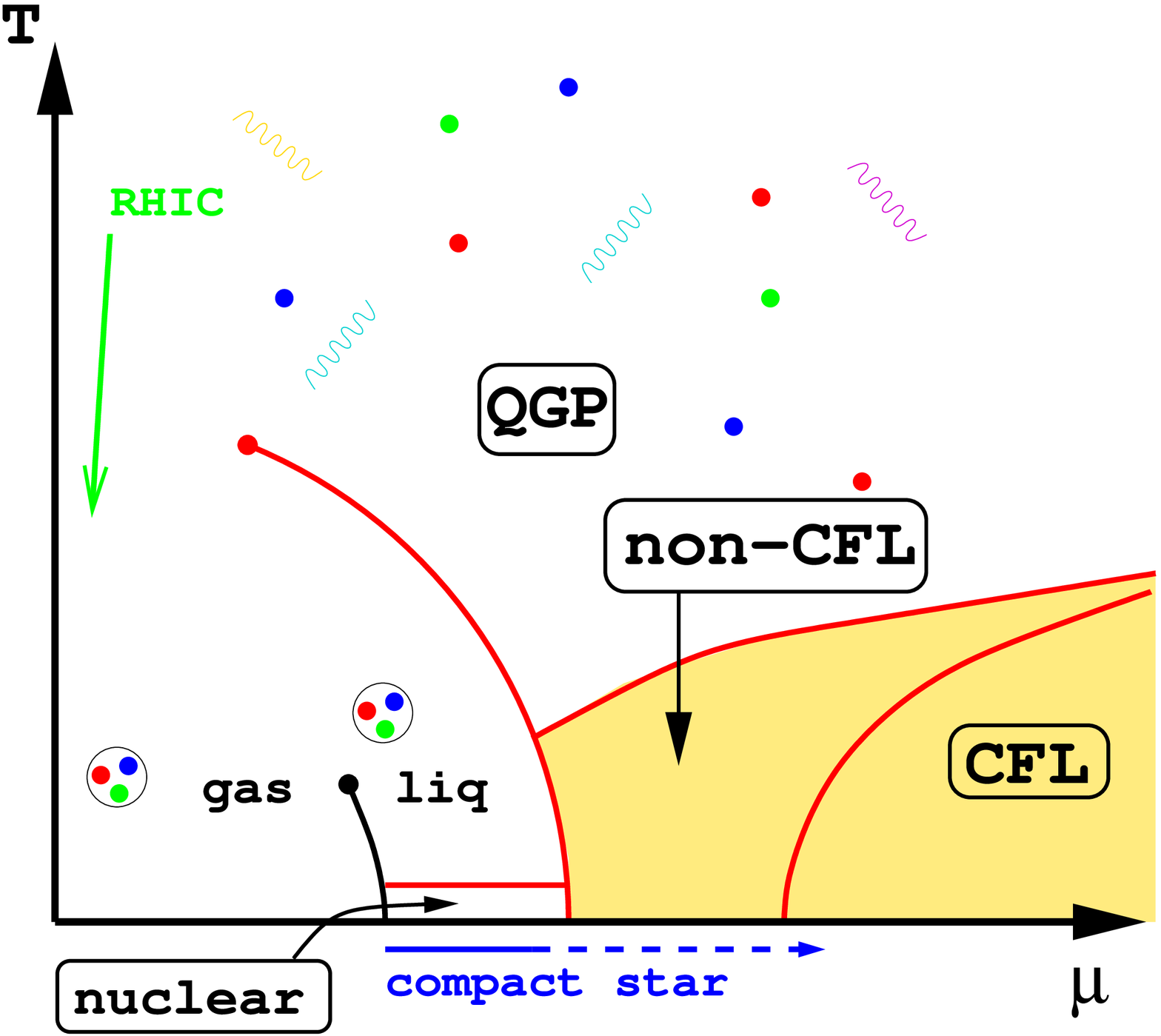}
}
\parbox{0.54\hsize}{
\begin{center} \underline{NJL calculation} \end{center}
 \includegraphics[width=\hsize]{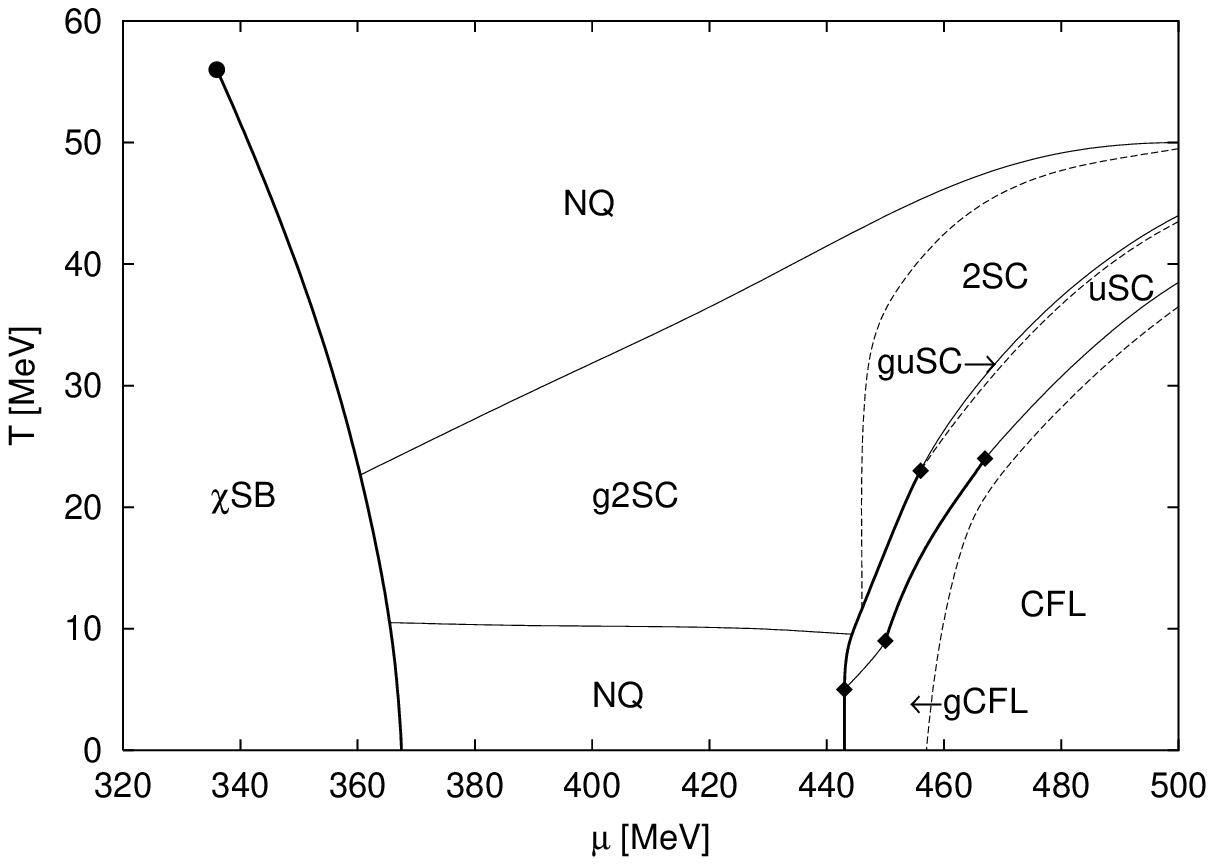}
}
\caption{On the left, the conjectured form of the phase diagram for matter
at ultra-high density and temperature.
On the right, the result of a calculation using a NJL model
\cite{Ruster:2005jc}.
At high density we find a rich structure of color-superconducting
phases.}
\label{fig:phase}
\end{figure}

In the real world there are two light quark flavors, the up 
($u$) and down ($d$), with 
masses $\lesssim 5~{\rm MeV}$, and a medium-weight flavor, the strange 
($s$) quark, with mass $\sim 100~{\rm MeV}$. (Their effective
``constituent'' masses in dense matter may be much larger.)
The strange quark therefore plays a crucial role in the phases of QCD.
Fig.~\ref{fig:phase} shows a conjectured phase diagram for QCD,
and also a calculated phase diagram obtained using a
Nambu--Jona-Lasinio model of QCD.
In both cases, along the horizontal axis the temperature is zero, and
the density rises from the onset of nuclear matter through the
transition to
quark matter. Compact stars are in this region of the phase diagram,
although it is not known whether their cores are dense enough
to reach the quark matter phase.
Along the vertical axis the temperature rises, taking
us through the crossover from a hadronic gas to the quark gluon plasma.
This is the regime explored by high-energy heavy-ion colliders.

At the highest densities we find the CFL phase, in which the strange
quark participates symmetrically with the up and down quarks
in Cooper pairing---this is described
in more detail below. The phases that occur at intermediate density
are still not well understood. The NJL calculation works with a
limited set of possibilities, and the NJL calculation
neglects more exotic possibilities such as kaon condensation
\cite{BedaqueSchaefer},
crystalline color superconductivity (LOFF) \cite{Alford:2000ze},
and single-flavor pairing 
\cite{IwaIwa,Schafer:2000tw,Buballa:2002wy,Alford:2002rz,Schmitt:2002sc}.

\subsection{Color superconductivity}

The fact that QCD is asymptotically free implies that
at sufficiently high density and low temperature,
there is a  Fermi surface of weakly-interacting quarks. 
The interaction between these quarks is
certainly attractive in some channels
(quarks bind together to form baryons), so we expect the formation
of a condensate of Cooper pairs.
We can see this by considering the grand canonical potential
$F= E-\mu N$, where $E$ is
the total energy of the system, $\mu$ is the chemical potential, and
$N$ is the number of quarks. The Fermi surface is defined by a
Fermi energy $E_F=\mu$, at which the free energy is minimized, so
adding or subtracting a single particle costs zero free energy. 
Now switch on a weak attractive interaction.
It costs no free energy to
add a pair of particles (or holes), and if they have the
right quantum numbers then the attractive
interaction between them will lower the free energy of the system.
Many such pairs will therefore
be created in the modes near the Fermi surface, and these pairs,
being bosonic, will form a condensate. The ground state will be a
superposition of states with all numbers of pairs, breaking the
fermion number symmetry. 

A pair of quarks cannot be a color singlet, so
the resulting condensate will break the local color symmetry
$SU(3)_{\rm color}$.  The formation of a condensate
of Cooper pairs of quarks is therefore called ``color superconductivity''.
The condensate plays the same role here as the Higgs particle
does in the standard model: the color-superconducting phase
can be thought of as the Higgs phase of QCD.

\vspace{\vpad}
\subsection{Highest density: Color-flavor locking (CFL)}
\label{sec:CFL}

At the highest densities, 
where
the strange quark Fermi momentum is close to the up and down
quark Fermi momenta, the
favored phase is ``color-flavor locking'' (CFL) \cite{CFL}.
This has been confirmed by both NJL \cite{CFL,Schafer:1999pb} and 
gluon-mediated interaction calculations \cite{Schafer:1999fe}.
The CFL pairing pattern is
\begin{equation}
\begin{array}{c}
\langle q^\alpha_i C \gamma_5 q^\beta_j \rangle^{\phantom\dagger}_{1PI}
\propto
  (\kappa+1)\delta^\alpha_i\delta^\beta_j 
+ (\kappa-1) \delta^\alpha_j\delta^\beta_i 
 = \eps^{\al\be N}\eps{ij N} + \ka(\cdots)  \\[2ex]
 {[SU(3)_{\rm color}]}
 \times \underbrace{SU(3)_L \times SU(3)_R}_{\displaystyle\supset [U(1)_Q]}
 \times U(1)_B 
 \to \underbrace{SU(3)_{C+L+R}}_{\displaystyle\supset [U(1)_{{\tilde Q} }]} 
  \times \mathbb{Z}_2
\end{array}
\end{equation}
Color indices $\alpha,\beta$ and flavor indices $i,j$ run from 1 to 3,
Dirac indices are suppressed,
and $C$ is the Dirac charge-conjugation matrix.
The term multiplied by $\kappa$ corresponds to pairing in the
$({\bf 6}_S,{\bf 6}_S)$, which
although not energetically favored
breaks no additional symmetries and so
$\kappa$ is in general small but not zero 
\cite{CFL,Schafer:1999fe,Shovkovy:1999mr,Pisarski:1999cn}.
The Kronecker deltas connect
color indices with flavor indices, so that the condensate is not
invariant under color rotations, nor under flavor rotations,
but only under simultaneous, equal and opposite, color and flavor
rotations. Since color is only a vector symmetry, this
condensate is only invariant under vector flavor+color rotations, and
breaks chiral symmetry. The features of the CFL pattern of condensation are
\begin{itemize}
\setlength{\itemsep}{-0.7\parsep}
\item[\Dash] The color gauge group is completely broken. All eight gluons
become massive. This ensures that there are no infrared divergences
associated with gluon propagators.
\item[\Dash]
All the quark modes are gapped. The nine quasiquarks 
(three colors times three flavors) fall into an ${\bf 8} \oplus {\bf 1}$
of the unbroken global $SU(3)$, so there are two
gap parameters. The singlet has a larger gap than the octet.
\item[\Dash] 
A rotated electromagnetism (``${\tilde Q} $'')
survives unbroken. It is a combination
of the original photon and one of the gluons.
\item[\Dash] Two global symmetries are broken,
the chiral symmetry and baryon number, so there are two 
gauge-invariant order parameters
that distinguish the CFL phase from the QGP,
and corresponding Goldstone bosons which are long-wavelength
disturbances of the order parameter. 
When the light quark mass is non-zero it explicitly breaks
the chiral symmetry and gives a mass
to the chiral Goldstone octet, but the CFL phase is still
a superfluid, distinguished by its baryon number breaking.
\item[\Dash]
The symmetries of the
3-flavor CFL phase are the same as those one might expect for 3-flavor
hypernuclear matter \cite{Schafer:1999pb}, so it is possible that there is
no phase transition between them.
\end{itemize}

\section{Real-world quark matter}

\subsection{Stresses on the CFL phase}
The CFL phase is characterized by pairing between different flavors
and different colors of quarks. We can easily understand why this
is to be expected. Firstly, the QCD interaction between two
quarks is most attractive in the channel that
is antisymmetric in color (the $\bar{\bf 3}$). Secondly, pairing
tends to be stronger in channels that do not break rotational symmetry
\cite{IwaIwa,Schafer:2000tw,Buballa:2002wy,Alford:2002rz,Schmitt:2002sc},
so we expect the pairing to be a spin singlet,
i.e.~antisymmetric in spin. Finally, 
fermionic antisymmetry of the Cooper pair wavefunction then
forces the Cooper pair to be antisymmetric in flavor.

Pairing between different colors/flavors can occur easily when
they all have the same chemical potentials and Fermi momenta.
This is the situation at very high density, where the strange quark
mass is negligible.
However, in a real compact star we must take into account the forces that
try to split those Fermi momenta apart, imposing an
energy cost on cross-species pairing. We must
require electromagnetic and color
neutrality \cite{Iida:2000ha,Alford:2002kj} (possibly
via mixing of oppositely-charged phases),
allow for equilibration under the weak interaction, and include a
realistic mass for the strange quark.  
These factors cause the different
colors and flavors to have different chemical potentials, and
this imposes a stress on cross-species pairing
such as occurs in the CFL pairing pattern.
As we come down in density, we expect the CFL pairing pattern
to be distorted, and then to be replaced by some other
pattern.

In the next few subsections we give a quick overview of the
expected phases of real-world quark matter. We restrict our
discussion to zero temperature because the critical temperatures
for most of the phases that we discuss are expected to be
of order $10~\MeV$ or higher, and the core temperature
of a neutron star is believed to drop below this value
within minutes (if not seconds) of its creation in a supernova.

%\vspace{\vpad}
\subsection{Kaon condensation: the CFL-$K^0$ phase}
Bedaque and Sch\"afer 
\cite{BedaqueSchaefer} showed that when the stress is not too large 
(high density), it may  simply
modify the CFL pairing pattern by inducing a flavor rotation of
the condensate which can be interpreted as a condensate
of ``$K^0$'' mesons, i.e.~the neutral anti-strange Goldstone bosons
associated with the chiral symmetry breaking. 
This is the ``CFL-K0'' phase, which breaks isospin.
The $K^0$ condensate can easily be suppressed by
instanton effects \cite{Schafer:2002ty}, but if these are
ignored then the kaon condensation occurs for
$M_s \gtrsim m^{1/3}\De^{2/3}$ for light ($u$ and $d$) quarks
of mass $m$.
% and $\mu^2 \lesssim \sqrt{m M_s}\De$. (??)
This was demonstrated
using an effective theory of the Goldstone bosons, but
with some effort can also be seen in an NJL calculation
\cite{Buballa:2004sx,Forbes:2004ww}.

\vspace{\vpad}
\subsection{The non-CFL region}
The nature of the next significant transition has been
studied in NJL model calculations which ignore the
$K0$-condensation in the CFL phase
\cite{Fukushima:2004zq,Blaschke:2005uj,Ruster:2005jc}.
It has been found that
the phase structure depends on the
strength of the pairing. If the pairing is very strong
(so that $\De_{CFL}\sim 100~\MeV$ where
$\De_{CFL}$ is what the CFL gap would be at $\mu\sim 500~\MeV$
if $M_s$ were zero)
then the CFL phase survives all the way down to the transition
to nuclear matter. For less strong pairing, there
may be a transition to a two-flavor pairing (``2SC'') phase
\cite{ARW2,RappEtc}
and/or a single-flavor pairing phase
(see below), and then to nuclear matter.

For a wide range of parameter values, however,
we find something
more interesting. We can make a
rough quantitative analysis by expanding in powers of
$M_s/\mu$ and $\De/\mu$, and
ignoring the fact that the effective strange
quark mass may be different in different phases \cite{Alford:2002kj}.
Such an analysis shows that as we come down in density 
we find a transition at $\mu \approx \half M_s^2/\De_{CFL}$
from CFL to another phase, the gapless CFL phase (gCFL) \cite{gCFL}.
The gCFL phase is also found in more complete NJL calcualtions
that do not use the assumptions of Ref \cite{gCFL}.
This is seen in the right-hand panel of Fig.~\ref{fig:phase}
(the ``gCFL'' region) and is discussed in detail below.
% subsections are not numbered in this package!!(sect.~\ref{sec:gCFL}).

\vspace{\vpad}
\subsection{Single-flavor pairing}
If $M_s$ is sufficiently large at densities where
quark matter is favored over nuclear matter, then
(via the neutrality requirement) it splits the chemical
potentials of the different flavors so far apart that
no cross-species pairing can occur at all. There is no
CFL or 2SC pairing. In most NJL studies this is described loosely
as ``unpaired'' quark matter. However, it is well known that
there are attractive channels for a single flavor, although they
are much weaker than the 2SC and CFL channels. Thus in these
regions we expect some form of single-flavor pairing.
There are various ``1SC'' phases, many of which
break rotational invariance, and a very interesting
color-spin-locked (CSL) phase which is rotationally invariant
\cite{IwaIwa,Schafer:2000tw,Buballa:2002wy,Alford:2002rz,Schmitt:2002sc}.
These phases have much lower critical temperatures than the others
(from a few MeV down to eV).

%\vspace{\vpad}
\subsection{The gapless CFL phase}
\label{sec:gCFL}

As mentioned above, an expansion
in $M_s/\mu$ for pairing strength $\De_{CFL}\lesssim 25~\MeV$
shows that,
for $M_s^2/\mu\gtrsim 2\De$, the CFL phase has higher free energy
than an alternative phase called gapless CFL (``gCFL'').
This follows from the energetic balance, mentioned above, between
the cost of keeping the Fermi surfaces together and the benefit
of the pairing that can then occur. 
The leading effect of $M_s$ is like a shift in the chemical potential
of the strange quarks, so the $bd$ and $gs$ quarks feel ``effective
chemical potentials''
$\mu_{bd}^{\rm eff} = \mu - \twothirds \mu_8$ and
$\mu_{gs}^{\rm eff} = \mu  + \third \mu_8 -\frac{M_s^2}{2\mu}$.
In the CFL phase $\mu_8=-M_s^2/(2\mu)$~\cite{Alford:2002kj},
%\beq
%\mu_{bd}^{\rm eff} - \mu_{gs}^{\rm eff} = \frac{M_s^2}{\mu}
%\label{mugsbd}
%\eeq
so $\mu_{bd}^{\rm eff} - \mu_{gs}^{\rm eff} = M_s^2/\mu$.
The CFL phase will be stable as long as the
pairing makes it energetically favorable to maintain equality of the
$bd$ and $gs$ Fermi momenta, despite their differing chemical
potentials \cite{CFLneutral}.
It becomes unstable when
the energy gained from turning a 
$gs$ quark near the common Fermi momentum into a $bd$ quark 
(namely $M_s^2/\mu$) exceeds the cost
in lost pairing energy $2\De_1$. 
So the CFL phase is stable when
\beq
\frac{M_s^2}{\mu} < 2\De_{CFL}\ ,
\label{CFLstable}
\eeq
For larger $M_s^2/\mu$, the CFL phase is      
replaced by some new phase with unpaired $bd$ quarks,
which  cannot be neutral unpaired
or 2SC quark matter because the 
new phase and the CFL phase must have the same 
free energy at the critical $M_s^2/\mu = 2\De_{CFL}$.

\begin{figure}[t]
 \bc
 \includegraphics[width=0.7\hsize]{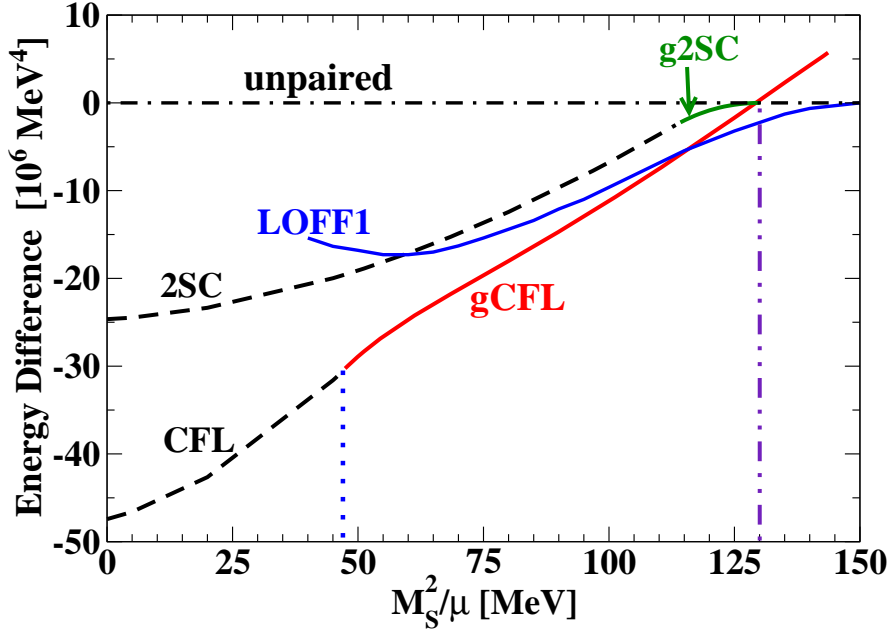}
 \ec
\caption{
Free energy of various phases in an NJL model, allowing
different pairing strengths  $\De_{ud}$, $\De_{ds}$, $\De_{us}$
for the different flavors. The CFL pairing strength is
$\De_{CFL}=25~\MeV$.
Note that the gCFL phase takes over
from CFL at $M_s^2/mu \approx 2 \De_{CFL}$, and remains favored
beyond the value $M_s^2/\mu \approx 4 \De_{CFL}$ at which the
CFL phase would become unfavored. The ``LOFF1'' curve is the
single-plane-wave LOFF ansatz of \cite{Casalbuoni:2005zp}.
}
\label{fig:energy}
\end{figure}

The obvious approach to finding this phase is to perform a NJL model
calculation with a general ansatz for the pairing that includes
differences between the flavors, for example by allowing different
pairing strengths $\De_{ud}$, $\De_{ds}$, $\De_{us}$. This was done
in Ref.~\cite{gCFL}, and the resultant ``gCFL'' phase was
described in detail. In Fig~\ref{fig:energy} we show the results.
The gCFL phase takes over
from CFL at $M_s^2/\mu \approx 2 \De_{CFL}$, and remains favored
beyond the value $M_s^2/\mu \approx 4 \De_{CFL}$ at which the
CFL phase would become unfavored.

\vspace{\vpad}
\subsection{Crystalline pairing}
The pairing patterns discussed so far have been
translationally invariant. But in the region of parameter space
where cross-species pairing is just barely excluded by stresses
that pull apart the Fermi surfaces, one expects a position-dependent
pairing known as the ``LOFF'' phase 
\cite{LOFF,Alford:2000ze,Bowers:2002xr,Casalbuoni:2003wh}.
This arises because one way
to achieve pairing between different flavors while accomodating the
tendency for the Fermi momenta to separate is to only pair over part
of the Fermi surface.
As we will discuss below, the LOFF phase
competes with the gCFL phase, and may resolve that phase's
stability problems.

\vspace{\vpad}
\subsection{Mixed Phases}
Another way for a system to deal with a stress on its
pairing pattern is phase separation. In the context of quark
matter this corresponds to relaxing the requirement of local charge
neutrality, and requiring neutrality only over long distances,
so we allow a mixture of a positively charged and a negatively charged
phase, with a common pressure and a common value of the electron
chemical potential $\mu_e$ that is not equal to the neutrality value
for either phase. Such a mixture of nuclear and CFL quark matter
was studied in Ref.~\cite{Alford:2001zr}. In quark matter
it has been found that as long as we require local color neutrality 
such mixed phases are not the favored response to the stress imposed
by the strange quark mass \cite{gCFL,Alford:2004nf}. Phases involving
color charge separation have been studied \cite{Neumann:2002jm} but it seems
likely that the energy cost of the color-electric fields will disfavor them.

\vspace{\vpad}
\subsection{Beyond gapless CFL}

\begin{figure}[ht]
 \bc
 \includegraphics[width=0.6\hsize]{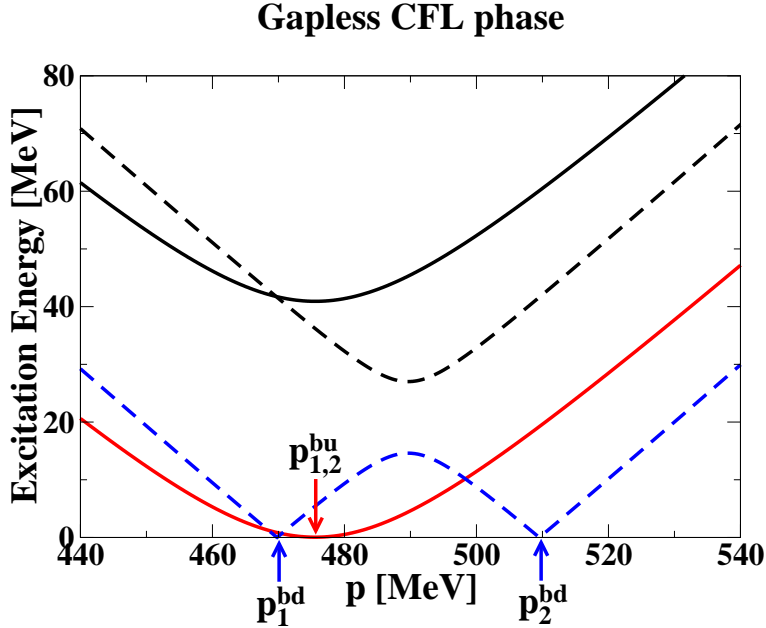}
 \ec
\caption{
Dispersion relations of the lightest quasiquark excitations
in the gCFL phase, at $\mu=500~\MeV$,
with $m_s=200~\MeV$ and  $\De_{CFL}=25~\MeV$.
Note that in there is a gapless mode with a 
{\em quadratic}\/ dispersion relation
(energy reaching zero at momentum $p^{bu}_{1,2}$) as well
as two gapless modes with more conventional linear dispersion
relations. 
}
\label{fig:disprel}
\end{figure}

The arguments above led us to the conclusion that the
favored phase of quark matter at the highest densities is the CFL
phase, and that as the density is decreased there is a transition
to another color superconducting phase, the gapless CFL phase.
However,  it turns out that the gCFL phase is itself unstable, 
and that there is therefore
another phase of even lower free energy, that occurs below the gCFL
phase in the phase diagram. The nature of that phase remains uncertain at 
present.

The instability of the gCFL phase appears to be related to one of its
most interesting features, namely
the presence of gapless fermionic excitations around the ground state.
These are illustrated in Fig.~\ref{fig:disprel}, which shows that
there is one mode (the $bu$-$rs$ quasiparticle) with
an unusual quadratic dispersion relation, which
is expected to give rise to exotic transport properties \cite{Alford:2004zr}.
The instability of the gCFL phase was established
in Refs.~\cite{Casalbuoni:2004tb,Fukushima:2005cm} after 
an analogous instability in the gapless 2SC phase had been discovered
\cite{Huang:2004am,Giannakis:2004pf}. The instability manifests itself
in imaginary Meissner masses $M_M$ for some of the gluons.
$M_M^2$ is the
low-momentum current-current two-point function, 
and $M_M^2/(e^2\De^2)$ is the 
coefficient of the 
gradient term in the effective theory of small fluctuations around the
ground-state condensate.
The fact that we find a negative value when the quasiparticles are gapless
indicates an instability 
towards spontaneous breaking of translational invariance.
Calculations in a simple two-species model \cite{Alford:2005qw}
show that imaginary $M_M$ is generically associated with
the presence of gapless charged fermionic modes.

The nature of the true ground state in the intermediate density regime
remains unclear. It could be
a mixed phase \cite{Reddy:2004my}, a crystalline (LOFF) phase
\cite{Giannakis:2004pf}, or a $p$-wave meson condensate 
\cite{Schafer:2005ym,Kryjevski:2005qq}.
Recent calculations \cite{Casalbuoni:2005zp} for the 3-flavor case show that
even a very simple LOFF ansatz yields a state 
that has lower free energy than gCFL
in the region where the gCFL$\to$unpaired transition occurs
(see Fig.~\ref{fig:energy}).
Based on what was found in the two-flavor case \cite{Bowers:2002xr},
it is reasonable to expect that when the full space of crystal
structures is explored, the LOFF state will be preferred to gCFL
over a much wider range of the stress parameter $M_s^2\De/(2\mu)$,
and it might turn out that the whole gCFL region is actually a LOFF region.

An alternative explanation was advanced by Hong \cite{Hong}
(see also Ref.~\cite{Huang:2003xd}): since the
instability is generically associated with the presence of gapless fermionic
modes, and the BCS mechanism implies that any gapless fermionic mode
is unstable to Cooper pairing in the most attractive channel, one might
expect that the instability will simply be resolved by ``secondary pairing''.
This means
the formation of a $\<qq\>$ condensate where $q$ is either one of the gapless
quasiparticles whose dispersion relation is shown in Fig.~\ref{fig:disprel}.
After the formation of such a secondary condensate, the linear 
gapless dispersion relations
would be modified by ``rounding out'' of the corner where the energy
falls to zero, leaving a secondary energy gap $\De_s$, which
renders the mode gapped, and removes the instability.
In the case of the quadratically gapless mode there is a greatly
increased density of states at low energy 
(in fact, the density of states diverges as $E^{-1/2}$), so Hong
calculated that the secondary pairing should
be much stronger than would be predicted by BCS theory, 
and he specifically predicted $\Delta_s \propto G_s^2$ 
for coupling strength $G_s$, as compared with
the standard BCS result
$\Delta \propto \exp(-{\rm const}/G)$. 

This possibility was worked out in an NJL model in
Ref.~\cite{Alford:2005kj}, using a two-species model. This allowed a
detailed exploration of the strength of secondary pairing. The
calculation confirmed Hong's prediction that in typical secondary
channels $\Delta_s \propto G_s^2$. However, in all the secondary
channels that were analyzed it was found that the secondary gap, even
with this enhancement, is from ten to hundreds of times smaller than
the primary gap at reasonable values of the secondary coupling. 
This shows that that secondary pairing
does not generically resolve the magnetic
instability of the gapless phase, since it indicates that there is
a temperature range $\De_s \ll T \ll \De_p$ in which there is primary
pairing (of strength $\De_p$)
but no secondary pairing, and at those temperatures the
instability problem would arise again.

\section{Conclusion}

As I have described, the project of delineating a plausible phase
diagram for high-density quark matter is still not complete. I have
discussed some ideas for the ``non-CFL'' region, but there are others
such as a suggested gluon condensation in two-flavor
quark matter \cite{Gorbar:2005rx}, and deformation of the Fermi surfaces
(discussed so far only in non-beta-equilibrated nuclear matter
\cite{Sedrakian:2003tr}).
It is very interesting to note that the problem of how a system with
pairing responds to a stress that separates the chemical potentials
of the pairing species is a very generic one, arising in
condensed matter systems and cold atom systems
as well as in quark matter.
Recent work
by Son and Stephanov \cite{Son:2005qx} on a two-species model characterized by
a diluteness parameter and a splitting potential
shows that between the BCS-paired region and the
unpaired region in the phase diagram 
one should expect a translationally-broken
region. In QCD this could correspond to a $p$-wave meson condensate
 or a LOFF state (see above). What is
particularly exciting is that the technology of cold atom traps
has advanced to the point where fermion superfluidity can
now be seen in conditions where many of the important parameters
can be manipulated, and it may soon be possible to investigate
the response of the pairing to external stress under
controlled experimental conditions.

%%%%%%%%%%%%%%%%%%%%%%%%%%%%%%%%%%%%%%%%%%%%%%%%
%% BACKMATTER
%%%%%%%%%%%%%%%%%%%%%%%%%%%%%%%%%%%%%%%%%%%%%%%%

%\begin{theacknowledgments}
%\end{theacknowledgments}

\end{document}